\documentclass[12pt]{article}

\usepackage{graphicx}

\title{Predicting the Ionization Threshold for Carriers in Excited Semiconductors}

\author{David Snoke\\Department of Physics and Astronomy \\ University of Pittsburgh, Pittsburgh, PA 15260}

\date{}

\begin{document}

\maketitle

\abstract{A simple set of formulas is presented which allows prediction of the fraction of ionized carriers in an electron-hole-exciton gas in a photoexcited semiconductor. These results are related to recent experiments with excitons in single and double quantum wells.}
\vspace{1cm}

Many researchers in semiconductor physics talk of ``the'' Mott transition density in a system of excitons and electron-hole plasma, but do not have a clear handle on exactly how to predict that density as a function of temperature and material parameters in a given system.  While numerical studies have been performed for the fraction of free carriers as a function of carrier density and temperature \cite{portnoi-2Dscreen,koch}, these do not give a readily-accessible intuition for the transition.  
In this paper I present a simple approach which does not involve heavy numerical methods, but is still fairly realistic.  The theory is based on two well-known approximations, which are the mass-action equation for equilibrium in when different species can form bound states, and the static (Debye) screening approximation. In addition, simple approximations are used for numerical calculations of the excitonic Rydberg as a function of screening length. 

\section{Two ``Mott'' transitions}

The first issue to deal with is what Mott transition we are concerned with. There are actually two very different conductor-insulator transitions which go under the name of the ``Mott'' transition. The first, originally envisioned by Mott, occurs when the wave function overlap of  bound states increases to the point that banding occurs. This Mott transition is relevant in the case of doping of semiconductors-- when the doping density increases beyond a certain point, a mini-band generated from the dopant states will arise, in which case the dopant carriers are no longer localized but instead can move in Bloch-like delocalized states. The condition for this is $r_s \sim a$, where $a$ is the characteristic length over which the bound state wave function extends, and $r_s$ is the average distance between bound states. If we cube both sides of this, we obtain the condition, for a three-dimensional system,
\begin{equation}
n \sim a^3,
\label{mott1}
\end{equation}
where $n$ is the density of dopant electrons.  

This Mott transition condition is equivalent to the condition that the Fermi level of the electrons, if they were not in bound states, is much larger than the binding energy in the bound states. We can see this by assuming hydrogenic bound states. In this case $a$ is given by the Bohr radius,
\begin{equation}
a = \frac{4\pi \epsilon \hbar^2}{e^2 m},
\end{equation}
where $\epsilon$ is the zero-frequency dielectric constant of the medium and $m$ is the effective mass of the bound carrier (MKS is used everywhere in this paper). The hydrogenic  Rydberg for the bound state is 
\begin{equation}
{\rm Ry} = \frac{e^2}{8\pi\epsilon a}.
\end{equation}
By comparison, the Fermi energy for a three-dimensional electron gas with two spin states, determined by the condition $N = \int_0^{E_F} D(E)dE$, where $D(E)$ is the density of states, is
\begin{equation}
E_F = \left[ n \frac{(3\pi^2)}{\sqrt{2}}\frac{\hbar^3 }{ 
m^{3/2} } \right]^{2/3} ,
\end{equation}
where $n$ is the electron density. 
If the Fermi energy is much larger than the Rydberg energy, e.g. 10 times greater, we
have
\begin{eqnarray}
\left(n
\frac{(3\pi^2)}{\sqrt{2}}\right)^{2/3}\frac{\hbar^2}{m}  &=& 
10\frac{e^2}{8\pi\epsilon a}\nonumber\\
2a\left(n\frac{(3\pi^2)}{\sqrt{2}}\right)^{2/3}\frac{4\pi\epsilon\hbar^2}{e^2m} 
&=&  10 \nonumber
\end{eqnarray}
or
$$
na^3  =  {10 \over 6\pi^2} \sim 1.
$$

Essentially, if the Fermi level is large compared to the binding energy in the bound states, the binding energy becomes just a perturbation on the Fermi gas. Since the Fermi gas is a conductor, the Coulomb binding energy has little effect on that. To put it another way, Fermi phase space filling effects, which lead to a Fermi level, become important when there is substantial wave function overlap of the electrons, which is the same point at which banding occurs. 

By contrast, there is another, {\em thermodnamic} insulator-conductor transition which is sometimes also called the Mott transition, although Rice has named it the ``ionization catastrophe.'' \cite{rice} This is usually the relevant transition for photoexcited carriers when the excited carrier density is low enough that there is no Fermi level, which is to say, the chemical potential is below the ground state energy. In this case, depending on the temperature, there can still be a conductor-insulator transition from the carriers mostly in excitons to the carriers mostly in plasma. The transition in this case comes about due to screening out of the Coulomb interaction which leads to formation of excitons, not due to wave function overlap.  The general condition is $a \sim l$, where $l$ is the screening length, which enters into the screened Coulomb (Yukawa) potential, $U(r) = q^2 e^{-r/l}/r$. The screening length depends on the density;  in the low-density limit in three dimensions it is given by the Debye screening formula (e.g. \cite{ridley}) 
\begin{equation}
\frac{1}{l^2} = \frac{e^2n}{\epsilon k_BT} .
\label{debye}
\end{equation}

We can envision this transition in two different directions. If we start with a fully ionized, high-density neutral plasma, and reduce the density while keeping the temperature constant (as, for example, typically happens due to carrier recombination) then an instability will occur when the total pair density drops to a point such that the screening length is long compared to the excitonic Bohr radius, and exciton formation is no longer screened out. Then essentially all the carriers can form into excitons. The condition for this is found by reversing the above formula, assuming $a$ is a constant:
\begin{equation}
n = \frac{ \epsilon k_BT}{a^2e^2}.
\label{nfree}
\end{equation} 

On the other hand, we can imagine starting with a low density gas of only excitons. Since excitons are neutral, we do not expect them to contribute to screening, at least not until their density is high enough that there is substantial wave function overlap, which is the condition (\ref{mott1}) above. However, long before they reach this density, free electrons and holes can be generated by thermal ionization. These free carriers can leading to screening, which can lead to a reduced exciton binding energy, which in turn leads to further ionization. At some critical density this can become a runaway process known as the ionization catastrophe, or a thermodynamic Mott transition.  To determine the condition for this, we must start with the mass-action equation, or Saha equation, for the number of free carriers in equilibrium,
\begin{equation}
\frac{n_e^2}{n_{ex} } = \frac{n_Q^{(e)} n_Q^{(h)}} {n_Q^{(ex)}} e^{-{\rm Ry}/k_BT} 
\label{massact}
\end{equation}
where $n_e$ is the density of free electrons (equal to the number of free holes), $n_{ex}$ is the density of excitons, and the other factors are the effective density of states factors, defined by
\begin{equation}
n_Q^{(i)} = \int_0^{\infty} D_i(E)e^{-E/k_BT}dE,
\end{equation}
where $D_i(E)$ is the density of states for species $i$, which depends in general on its band mass $m_i$ and spin degeneracy.  The mass-action equation follows straightforwardly from the equilibrium condition that the chemical potentials for the different species in a chemical reaction must balance, that is, $\mu_e + \mu_h = \mu_{ex}$, and from the Maxwellian (low-density) formula for the total density,
\begin{equation}
n_i = e^{-(E_i-\mu)/k_BT}n_Q^{(i)} ,
\end{equation}
where $E_i$ is the ground state energy of species $i$, and from the definition of the binding energy ${\rm Ry} = (E_c - E_v) - E_{ex}$.  Number conservation requires that $n_{ex} + n_e = n$, where $n$ is the total pair density at any point in time.

To find the critical density, we set the {\em free} carrier density $n_e$ to the value (\ref{nfree}) above, instead of the total density. Above this density, screening will make exciton formation impossible. 
Assuming ${\rm Ry}$ is constant and $n_e \ll n_{ex}$,  so that $n \approx n_{ex}$, this means the
critical density for the transition is approximately
\begin{equation}
n   = \left(\frac{ \epsilon k_BT}{a^2e^2}\right)^2\frac{e^{{\rm Ry}/k_BT}}{n_Q},
\label{ioncat}
\end{equation}
where $n_Q = n_Q^{(e)} n_Q^{(h)}/ n_Q^{(ex)}$.
This critical density, for starting at low density and going up, clearly gives a very different threshold than (\ref{nfree}) above, for starting at high density and going down. For example, at low temperature, very few excitons will ionize and therefore the critical density becomes exponentially large. Of course, at some point the Fermi-level Mott transition density (\ref{mott1}) will be reached. In general, though, the density predicted by (\ref{ioncat}) is well below the density predicted by (\ref{mott1}).

The difference between (\ref{nfree}) and (\ref{ioncat}) depending on whether the system is starting with excitons or neutral plasma leads to the possibility of {\em hysteresis}, that is, the possibility that the ionized fraction of excitons can depend not only on the total excitation density and temperature of the carrier gas, but also on its history. This possibility was first presented in Ref.~\cite{craw} and has been recently revived by Portnoi \cite{portnoi} for two-dimensional systems; experimental evidence has been reported in Ref.~\cite{crawexp}.

As seen in this short survey, talking of the Mott transition density is nontrivial.  The conductor-insulator transition in a carrier gas depends on density, temperature, the depth of the bound states, and the history of the system.
 
 \section{Self-consistent calculations for a three-dimensional system}
 
 In this section we ignore the possibility of hysteresis and ask for the prediction of the mass-action equation for the fraction of ionized carriers as a function of density and temperature. If we are to be self-consistent, we must account for the fact that the binding energy used in the mass-action equation (\ref{massact}) can depend on density through the density-dependent screening length. A very useful approximation, based on numerical calculations \cite{smith}, is
  \begin{equation}
{\rm Ry}(l) =  \left\{ \begin{array}{ll} \displaystyle
{\rm Ry}(0) \left( 1 - \frac{2}{1+l/a}\right), &  l/a>1\\
0, & l/a \le 1, \end{array} \right.
\label{smith}
\end{equation}
where $l$ is the screening length. Using this formula, and equation (\ref{debye}) for the screening length, we can rewrite (\ref{massact}) using number conservation as
\begin{equation}
n_e^2 = (n-n_e)n_Qe^{-{\rm Ry}(n_e,T)/k_BT},
\end{equation}
which can be solved for $n_e$. 

It turns out that we must adjust this equation slightly. The mass-action equation (\ref{massact}) breaks down in the limit ${\rm Ry} \rightarrow 0$, which is exactly the limit we are interested in for the ionization catastrophe. This is easily seen by solving for $n_{ex}$ when ${\rm Ry} = 0$, which gives $n_{ex} = n_e^2/n_Q$, which is not zero. We expect, however, that the number of excitons vanishes when the exciton binding energy is zero. The reason for the breakdown is that exciton states with extremely shallow binding are still counted as exciton states, when in reality such states would be unstable, with very large orbits, and electrons and holes in such orbits would be indistinguishable in practice from free electrons and holes. 
As a first-order correction, one can subtract from $n_{ex}$ the fraction of excitons which are in states with energy greater than ${\rm Ry}$, namely a fraction equal to $e^{-{\rm
Ry}/k_BT}$, in which case the mass-action equation becomes
\begin{equation}
\frac{n_en_h}{n_{ex} } = \frac{n_Q^{(e)} n_Q^{(h)}e^{-{\rm
Ry}/k_BT}} {n_Q^{(ex)}(1-e^{-{\rm
Ry}/k_BT})}  =  \frac{n_Q}{e^{{\rm
Ry}/k_BT}-1} ,
\label{massactadj}
\end{equation}
which has the expected behavior that $n_{ex} \rightarrow 0$ when ${\rm Ry} \rightarrow 0$. 
Much theoretical effort has been given, especially in the astrophysics community, to justifying this approach more rigorously, for example using Planck-Larkin partition functions, which give higher-order corrections \cite{ebeling,kraft}. 

Using this adusted mass-action equation, we then have
\begin{equation}
n_e^2 + n_e \left(\frac{n_Q}{e^{{\rm Ry}(n_e,T)/k_BT}-1}\right) - \frac{nn_Q}{e^{{\rm Ry}(n_e,T)/k_BT}-1} = 0.
\label{3deq}
\end{equation}
If ${\rm Ry}(n_e,T)$ is taken as a constant, then this equation can be solved simply as a quadratic equation. This leads to a value for $n_e$ which can then be used to recompute the value of ${\rm Ry}(n_e,T)$. This leads to an iterative process to converge on a self-consistent value of $n_e$.  This iteration converges very rapidly, typically in less than 10 iterations.  These formulas can be simply programmed in a conditional loop to generate the curves shown in this paper.

\begin{figure}
\begin{center}
\includegraphics[width=0.75\textwidth]{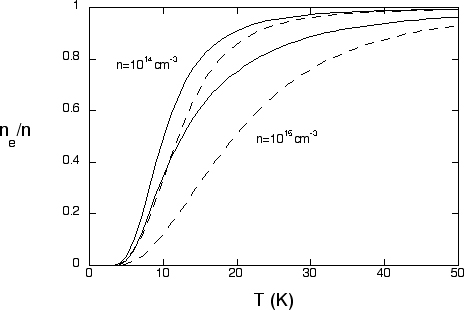}
\end{center}
\caption{Solid lines: fraction of free carriers in bulk GaAs predicted by Equation (\protect\ref{3deq}) in the text, as a function of temperature for the two total pair densities labeled. Parameters used are $m_e = 0.07 m_0$, $m_h = 0.08 m_0$, $\epsilon/\epsilon_0 = 12$, and ${\rm Ry}(0) = 4$ meV, which implies $a = 150$~\AA. Dashed lines: the prediction for constant excitonic Rydberg Ry $= 4$ meV, for the same two densities.}
\end{figure} 
 
Figure 1 shows the fraction of free carriers in bulk GaAs, with the expected behavior that the fraction increases toward 100\% as temperature is increased and the excitons ionize.  For simplicity we assume that all holes have the light hole mass; the presence of heavy holes will give a correction but not a substantial change to the forms of the curves. The dashed curves show the prediction of the mass-action equation for the same parameters, but constant excitonic Rydberg.  As expected, the self-consistent Rydberg solution gives a small correction at low density, and a larger correction at higher density. The self-consistent prediction becomes the same as the constant-Rydberg prediction at low temperature; for the parameters used in Fig.~1 the two predictions are identical below 4 K. 

We can also plot the ionized fraction as a function of density at fixed temperature. Figure 2 shows the fraction of ionized carriers as a function of density for various temperatures predicted for bulk GaAs.  At low temperature, the insulator-conductor transition is quite sharp, while at higher temperature it is less so, but at all temperatures there is a ``catastrophe'' at which $n_e/n$ becomes strictly equal to unity. Below $T= 4$ K, the critical density predicted by these equations becomes greater than that of the phase-space filling Mott transition density, $n \simeq 1/a^3 = 3\times 10^{17}$ cm$^{-3}$.  Surprisingly, the mass-action also predicts that at extremely low densities, the fraction of free carriers  approaches 100\% at all temperatures. This is because the rate of formation of excitons is proportional to $n_e^2$, and at low density the electrons simply cannot find each other to form excitons. 

\begin{figure}
\begin{center}
\includegraphics[width=0.75\textwidth]{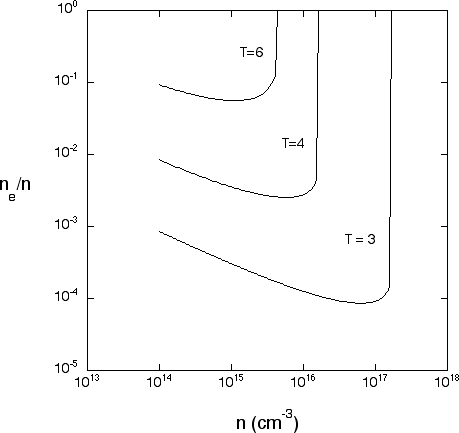}
\end{center}
\caption{Fraction of free carriers in bulk GaAs predicted by Equation (\protect\ref{3deq}) in the text, as a function of density for the four temperatures shown.  Parameters for bulk GaAs are the same as for Fig. 1. }
\end{figure} 

Figure 3 shows the predicted phase boundary. At low temperatures, this is a sharp conductor-insulator phase boundary. As seen in Fig. 1, above around 6 K a substantial fraction of free carriers coexist with excitons even below the ionization catastrophe boundary.  The effect of the transition will be much more dramatic on the luminescence spectrum, however, since excitons dominate the luminescence due to their stronger oscillator strength. At the ionization catastrophe, the exciton line will disappear.

 \begin{figure}
 \begin{center}
\includegraphics[width=0.75\textwidth]{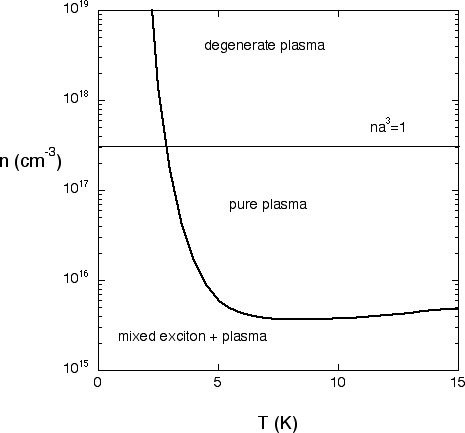}
\end{center}
\caption{Predicted phase diagram of carriers in bulk GaAs. Heavy line: the critical threshold for 100\% ionization implied by Equation \protect\ref{3deq} in the text. Above the boundary, the system is a pure plasma. Light line: the phase space filling condition $ n = 1/a^3 \sim 3\times 10^{17}$ cm$^{-3}$.  Parameters for bulk GaAs are the same as for Figs. 1 and 2. }
\end{figure} 

\section{Calculations for a two-dimensional System}
 
We can use the same approach for a two-dimensional system, e.g. carriers in quantum wells.  In this case the screening length is given by \cite{banyai}
\begin{equation}
\frac{1}{l} = \frac{e^2n}{2\epsilon k_BT} .
\label{debye2}
\end{equation}
(Note that the Coulomb interaction is still assumed to be three-dimensional, but the carriers are constrained to move only in two dimensions.) The two-dimensional exciton Bohr radius is related to the two-dimensional excitonic Rydberg by \cite{haugkoch}
\begin{equation}
{\rm Ry} = \frac{e^2}{4\pi\epsilon a}. 
\end{equation}
In addition, we must use two-dimensional densities of states.  A numerical solution for the variation of the Rydberg as screening increases in two dimensions has been performed by Portnoi and Galbraith \cite{portnoi-2Dscreen}, which can be approximated as 
  \begin{equation}
{\rm Ry}(l) =  \left\{ \begin{array}{ll} \displaystyle
{\rm Ry}(0) \left( 1 - \frac{2}{1+\sqrt{2l/a}}\right), &  l/a>\frac{1}{2}\\
0, & l/a \le \frac{1}{2}, \end{array} \right.
\end{equation}
where $a$ is the 2D excitonic Bohr radius.

\begin{figure}
\begin{center}
\includegraphics[width=0.75\textwidth]{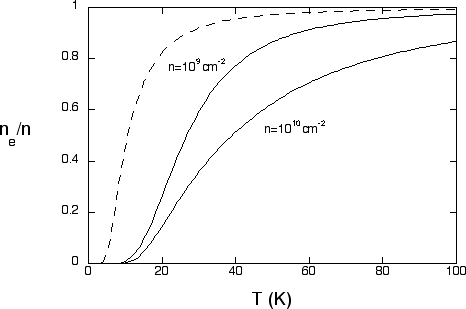}
\end{center}
\caption{Fraction of free carriers in single GaAs quantum wells predicted by Equation \protect\ref{3deq} in the text, as a function of temperature for the densities shown.  Solid lines: parameters for GaAs are the same as for Fig. 1, but ${\rm Ry} = 10$ meV $a = 120$ \AA, typical values for an exciton in a narrow single quantum well. Dashed line: the same parameters, but ${\rm Ry} = 4$ meV, for the lower density.}
\end{figure} 

Figure 4 shows the fraction of free carriers as a function of temperature for the parameters of narrow GaAs quantum wells, with ${\rm Ry} = 10$ meV,  for two densities. In this case only one in-plane mass is relevant at low temperature.  This figure also shows a curve for the same parameters but ${\rm Ry} = 4$ meV, with the same exciton Bohr radius, which are realistic parameters for indirect excitons in double quantum wells, which have been used in recent experiments \cite{rings,szy}.     It is perhaps surprising that excitons exist all the way up to 100 K or above at these densities.  This is a simple consequence of the prediction of the mass-action equation (\ref{massactadj}), which in the limit of low density and high temperature is
\begin{equation}
\frac{n_{ex}}{n} = \frac{n}{n_Q}\frac{\Delta}{k_BT}.
\end{equation}
Although higher temperature favors ionization, the free carrier fraction is never 100\% unless there is an ionization catastrophe.  
The fraction of the gas which is excitons is very low at these high temperatures, (e.g. about 1\% at $T=90$ K for 4 meV binding energy), but since they have a strong oscillator strength they can still have a strong emission in the luminescence in many semiconductors. 

Figure 5 shows the fraction of free carriers as a function of total pair density for several temperatures, for the case of single quantum wells. A sharp transition occurs at all temperatures, above which $n_e/n$ is exactly equal to unity. The behavior is similar to that of bulk, but with the temperature scale shifted to higher temperature and density. Figure 6 shows the predicted phase diagram, for the two different values of exciton binding energy discussed above.  

 \begin{figure}
 \begin{center}
\includegraphics[width=0.75\textwidth]{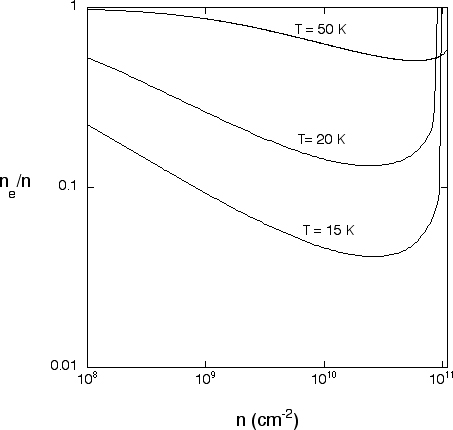}
\end{center}
\caption{Fraction of free carriers in single GaAs quantum wells predicted by Equation \protect\ref{3deq} in the text, as a function of density for the temperatures shown.  Parameters for GaAs are the same as for Fig. 4. }
\end{figure} 

 \begin{figure}
 \begin{center}
\includegraphics[width=0.75\textwidth]{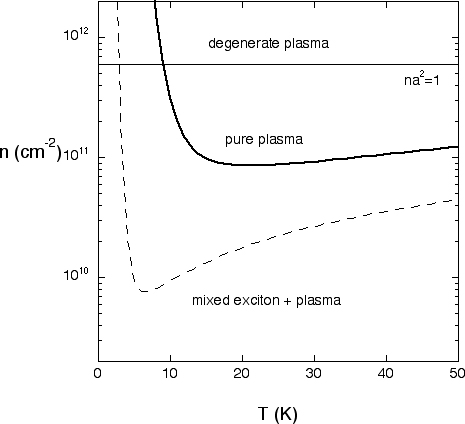}
\end{center}
\caption{Predicted phase diagram of carriers in GaAs quantum wells, for two different exciton binding energies. (Heavy solid line: ${\rm Ry} = 10$ meV ; dashed line: ${\rm Ry} = 4$ meV .)  Heavy line: the critical threshold for 100\% ionization according to the model discussed in the text. Above the boundary, the system is a pure plasma. Light line: the phase space filling condition $ n = 1/a^2$. }
\end{figure} 
 
\section{Conclusions}

The self-consistent mass action equation presented here allows a relatively simple way of calculating the fraction of excitons in equilibrium in a population of excited carriers, and also the critical density for the thermodynamic Mott transition, or ionization catastrophe, in which all excitons are screened out. 

Although the phase boundaries for the thermodynamic Mott transition are not simple, their form allows us to speak of two critical thresholds. At moderate densities (within an order of magnitude or so of the phase space filling density $n = 1/a^d$, where $d$ is the dimension), one can speak of a critical temperature for the Mott transition, since the phase boundary corresponds to roughly constant temperature in this range.  On the other hand, at high temperature, one can speak of a critical density for the Mott transition, since the phase boundary corresponds to nearly constant density in this region.  This critical density is typically an order of magnitude or more below the phase space filling density $n = 1/a^d$ in both two and three dimensions. 
 
At moderate densities, the critical temperature for the Mott transition is much less than ${\rm Ry}/k_B$.  On the other hand, at low densities, excitons persist as a small fraction of the gas to temperatures well above ${\rm Ry}/k_B$.

\end{document}